\documentclass[%
    reprint,
    superscriptaddress,
     amsmath,amssymb,
     aps,
    floatfix,
    ]{revtex4-2}
    
    \usepackage{graphicx}% Include figure files
    \usepackage{dcolumn}% Align table columns on decimal point
    \usepackage{bm}% bold math
    \usepackage{placeins}%N added for fixing the pic and table
    \usepackage{afterpage}
    \usepackage{layouts}
    \usepackage{xcolor}
    \usepackage{textgreek}
\begin{document}    
\title{The ideal wavelength for daylight free-space quantum key distribution}
    
\author{Mostafa Abasifard}
\email{mostafa.abasifard@uni-jena.de}
\affiliation{Abbe Center of Photonics, Institute of Applied Physics, Friedrich Schiller University Jena, 07745 Jena, Germany}
\author{Chanaprom Cholsuk}
\affiliation{Abbe Center of Photonics, Institute of Applied Physics, Friedrich Schiller University Jena, 07745 Jena, Germany}
\author{Roberto G. Pousa}
\affiliation{Computational Nonlinear and Quantum Optics, SUPA Department of Physics, University of Strathclyde, Glasgow G4 0NG, United Kingdom}
\author{Anand Kumar}
\affiliation{Abbe Center of Photonics, Institute of Applied Physics, Friedrich Schiller University Jena, 07745 Jena, Germany}
\author{Ashkan Zand}
\affiliation{Abbe Center of Photonics, Institute of Applied Physics, Friedrich Schiller University Jena, 07745 Jena, Germany}
\author{Thomas Riel}
\affiliation{University of Applied Sciences Wiener Neustadt, 2700 Wiener Neustadt, Austria}
\author{Daniel K. L. Oi}
\affiliation{Computational Nonlinear and Quantum Optics, SUPA Department of Physics, University of Strathclyde, Glasgow G4 0NG, United Kingdom}
\affiliation{Walton Institute for Information and Communication Systems Science, South East Technological University, Waterford X91 P20H, Ireland}
\author{Tobias Vogl}
\email{tobias.vogl@tum.de}
\affiliation{Abbe Center of Photonics, Institute of Applied Physics, Friedrich Schiller University Jena, 07745 Jena, Germany}
\affiliation{Department of Computer Engineering, School of Computation, Information and Technology, Technical University of Munich, 80333 Munich, Germany}

\begin{abstract}
Quantum key distribution (QKD) has matured in recent years from laboratory proof-of-principle demonstrations to commercially available systems. One of the major bottlenecks is the limited communication distance in fiber networks due to the exponential signal damping. To bridge intercontinental distances, low Earth orbit satellites transmitting the quantum signals over the atmosphere can be used. These free-space links, however, can only operate during the night, as the sunlight otherwise saturates the detectors used to measure the quantum states. For applying QKD in a global quantum internet with continuous availability and high data rates, operation during daylight is required. In this work, we model a satellite-to-ground quantum channel for different quantum light sources to identify the optimal wavelength for free-space QKD in ambient conditions. Daylight quantum communication is possible within the Fraunhofer lines or in the near-infrared spectrum, where the intrinsic background from the sun is comparably low. The highest annual secret key length considering the finite key effect is achievable at the H\textalpha\ Fraunhofer line. More importantly, we provide the full model that can be adapted in general to any other specific link scenario. We also propose a true single-photon source based on a color center in hexagonal boron nitride coupled to a microresonator that can implement such a scheme. Our results can also be applied in roof-to-roof scenarios and are therefore relevant for near-future quantum networks.
\end{abstract}

\maketitle

\section{Introduction}
Quantum key distribution (QKD) offers a level of security for cryptography that is bounded only by the fundamental laws of quantum physics and is therefore independent of any resources available for cryptanalysis \cite{gisin2002quantum,scarani2009security,pirandola2020advances}. As it is generally believed that quantum physical laws are invariant in time, QKD features perfect forward and backward secrecy. This unique opportunity for ultimate security and privacy has resulted in great interest over the past decades. Point-to-point quantum communication has been realized over distances ranging from a few meters for mobile applications \cite{vest2022quantum} up to hundreds of kilometers with optical fibers \cite{gobby2004quantum,boaron2018secure,chen2021twin}. In addition, local metropolitan and transnational networks have been established as proof-of-principle \cite{peev2009secoqc, sasaki2011field, chen2021implementation,Neumann2022}. The exponential scattering and absorption losses in fibers, however, prevent building a global quantum internet.\\
\indent The current distance record over fiber is 1002 km with a secret data rate of 0.003 bps using a twin-field QKD protocol \cite{PhysRevLett.130.210801}, surpassing the previous record at 830 km \cite{wang2022twin}. The losses for twin-field QKD protocols enter with the square root into the rate-distance relationship (unlike conventional protocols), however, the scaling remains exponential. Overcoming global distances and maintaining sufficiently high data rates require satellite links \cite{bedington2017progress} or quantum repeaters \cite{briegel1998quantum,vinay2017practical,zwerger2018long,Gundogan2021}. The latter will not be available in the near future, due to the technical challenges of building efficient quantum repeaters and quantum memories. Given that the attenuation in the atmosphere becomes negligible above 10 km \cite{peng2005experimental}, satellite-mediated QKD provides a promising alternative to quantum repeaters \cite{vallone2015experimental,sidhu2021advances,cao2022evolution}. The first space-to-ground quantum key exchange has been demonstrated by the Micius satellite achieving an average rate of 1.1 kbps during the satellite passage using the BB84 protocol with decoy states \cite{liao2017satellite,lu2022micius}. The link to Micius was only operational during the night to avoid sunlight saturating the single-photon detectors in the ground station.\\
\indent Near future quantum networks will require daylight operation for continuous availability. Several studies have explored daytime QKD using weak coherent states produced by pulsed lasers \cite{buttler2000daylight, hughes2002practical, peloso2009daylight, ko2018experimental, liao2017long, gong2018free, avesani2021full}. The wavelengths were chosen in the atmospheric transmission windows in the range of 700-900 nm or in the telecom C-band at 1550 nm \cite{liao2017long, gong2018free, avesani2021full}. The latter has the appeal of intrinsically lower solar irradiance, at the expense of less efficient single-photon avalanche diodes (SPADs) based on e.g.\ InGaAs/InP (compared to silicon-based SPADs) or very expensive (but efficient) superconducting nanowire single-photon detectors (SNSPDs). Another major drawback when using longer wavelengths is the linear scaling of the divergence with the wavelength. The diffraction losses can be compensated by larger telescopes, however, this drastically increases the cost of satellites and optical ground stations. Quantum communication using visible wavelengths is therefore desirable.\\
\indent The solar background can be suppressed by spectral, spatial, and temporal filtering. A unique opportunity is offered by the Fraunhofer lines, the set of dark lines in the solar spectrum caused by absorption in the Sun's or the Earth's atmosphere. When quantum information is encoded in light at these wavelengths, it can be separated from the sunlight by spectral filtering in the receiver. This scheme was originally proposed in 2006 \cite{rogers2006free} but never actually realized due to the need for a tunable short pulse laser and either a fast polarization modulator for visible light or multiple lasers that are spatially overlapped. This initial work proposed to use the first hydrogen transition of the Balmer series (H\textalpha) at a wavelength of 656.448 nm, as this is the broadest Fraunhofer line. The question arises, which wavelength in general is the most efficient for QKD for enabling daylight quantum communication, and which realistic light source can realize such a scheme?\\
\indent Moreover, within the context of satellite QKD, the finite key effect stands out as an influential factor \cite{sidhu2022finite}. Its significance is amplified due to the restrictive transmission times between satellite and ground station, which impose a limit on the generation of secret keys. This constraint is primarily influenced by two intertwined factors. The first is that the conventional asymptotic resource assumption tends to fall short as an effective approximation when dealing with short received signal blocks. This shortcoming arises from the absence of an arbitrarily large number of received signals, making it impossible to overlook statistical uncertainties. As a result, ensuring the security of the distilled secret key demands meticulous treatment of the statistical fluctuations inherent in the estimated parameters. The second consideration is the increasing need for optimizing the balance between the proportion of signals allocated for parameter estimation and those employed for key generation. In the aftermath of post-processing operations, including error correction, the amount of extractable secret key is diminished. This situation is further complicated by small block lengths, which induce additional inefficiencies surpassing the asymptotic limit.\\
\indent In this work, we identify the ideal wavelength for daylight free-space QKD for satellite-to-ground scenarios. We examine wavelengths ranging from 400 to 1700 nm and obtain the extractable secret bit per signal as the performance metric for each wavelength. To calculate this, we take the solar background spectrum and the transmission spectrum through the atmosphere and assume realistic parameters for telescope sizes, detection efficiencies and dark count rates from commercially available single-photon detectors, as well as different types of quantum light sources. The latter includes an ideal single-photon source (SPS), a suitable (realistic) single-photon source (`our SPS'), a weak coherent source (WCS), and a two-decoy state source (TDS). The realistic single-photon source assumes the parameters from a tunable microcavity-coupled color center in hexagonal boron nitride (hBN) \cite{nnano.2015.242,10.1021/acsphotonics.8b00127} that we have developed in past work \cite{vogl2019compact}. This true single-photon source operates at room temperature and has been implemented on a satellite platform, hence, it is suitable for the here investigated space-to-ground scenarios. We integrate over an entire satellite pass to calculate the total secret information and consider daylight and nighttime conditions. We aim to investigate the performance of different wavelengths and light sources. Therefore, we start with the asymptotic model and then perform finite secret key length analysis to utilize the tight finite block statistics in conjunction with system configuration and functional considerations.
\section{\label{sec:methods}Asymptotic Model}
We first provide the models that we used to calculate the spectrum of the secret key rates. The following section, therefore, describes a realistic QKD system taking into account the sender, quantum channel, and receiver unit for the space-to-ground quantum communication link.
\subsection{Geometric considerations}
    \begin{figure}
    \includegraphics[width=8.5cm]{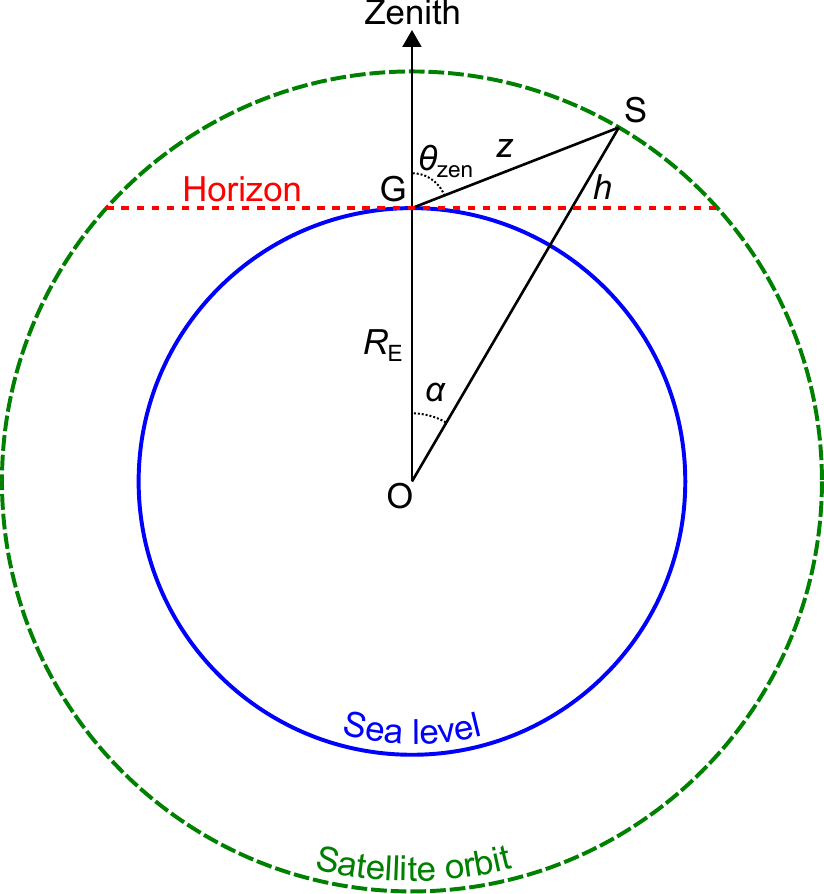}
    \caption{The geometry of the satellite orbit around the Earth and parameter definition. The ground station is assumed to be at sea level ($h_0=0$). Note that the horizon line is the astronomical horizon, not the true horizon.}
    \label{fig:geometry}
    \end{figure} 
We consider an optical ground station (G) at sea level and a satellite (S) orbiting at some variable altitude $h$ above the K\'arm\'an line that defines the boundary between the Earth's atmosphere and outer space (100 km above sea level). The angle between the zenith and the direct line of sight (G-S) is $\theta_\mathrm{zen}$ (see Fig.\ \ref{fig:geometry}). It takes values from $-\pi/2$ (front horizon) to $+\pi/2$ (back horizon). For a more general case, the ground station can be located at some nonzero altitude $h_0$. The slant distance $z$ between the satellite and the ground station can be calculated from simple geometric considerations (see Fig.\ \ref{fig:geometry}) and is given by
\begin{equation}
z(h, \theta_\mathrm{zen})= \sqrt{R_\mathrm{S}^2 + R_\mathrm{G}^2(\cos^2 \theta_\mathrm{zen} -1)} - R_\mathrm{G}\cos \theta_\mathrm{zen},
\end{equation}
where $R_\mathrm{G}= \mathrm{R_E}+h_0$ ($R_\mathrm{S}= \mathrm{R_E}+h$) is the distance of the Earth's center and the ground station (satellite). The radius of the Earth is $\mathrm{R_E}= 6371$ km when imagined as a perfect sphere with the global average value. Equivalently, the altitude $h$ of the satellite reads
\begin{equation}
h(z, \theta_\mathrm{zen})= \sqrt{R_\mathrm{G}^2+ z^2 +2z R_\mathrm{G} \cos \theta_\mathrm{zen}} - \mathrm{R_E}.
\end{equation}
For simplicity, we restrict the following analysis to zenith-crossing circular orbits (our results can be generalized to any arbitrary orbit). The slant distance $z=z(R_\mathrm{S}, \alpha)$ can be formulated as $z(R_\mathrm{S}, \alpha)=\sqrt{R_\mathrm{G}^2+R_\mathrm{S}^2-2R_\mathrm{G} R_\mathrm{S} \cos  \alpha}$ in terms of $R_\mathrm{S}$ and the orbital angle $\alpha$ (see Fig.\ \ref{fig:geometry}). Subsequently, we can define the orbital period as (i.e., the time needed for a full orbit around the Earth) as
\begin{equation}
T_\mathrm{S} = 2 \pi \sqrt{\frac{R_\mathrm{S}^3}{\mathrm{G}\mathrm{M_E}}},
\end{equation}
where $\mathrm{G}=6.674\times10^{-11} \mathrm{N m^2 kg^{-2}}$ is the gravitational constant and $\mathrm{M_E} = 5.972 \times 10^{24}$ kg is the Earth's mass. This parameterization is useful for circular orbits, where $R_\mathrm{S}$ is constant. In this case, we can write $\alpha=2\pi t/T_\mathrm{S}$. The orbital time can be expressed by
\begin{equation}
t(\theta_\mathrm{zen}, h)  = \sqrt{\frac{(\mathrm{R_E}+h)^3}{\mathrm{G}\mathrm{M_E}}} \arccos \left[\frac{R_\mathrm{G}+z(h, \theta_\mathrm{zen}) \cos \theta_\mathrm{zen}}{\mathrm{R_E}+h}\right],
\end{equation}
assuming $\theta_\mathrm{zen}=0$ for $t=0$. It might be convenient to shift the time such that the satellite crosses the horizon at $t=0$ to not have negative values for the time axis. For the altitude of the ground station, we assume sea level ($h_0=0$) in the following.
\subsection{Diffraction-induced transmissivity}
We assume that free-space quantum communication is based on a quasi-monochromatic optical mode represented by a Gaussian beam with waist $w_0$ and infinitely large curvature. The spot size is matching the telescope's clear aperture such that the aperture is not inducing any diffraction. After free-space propagation for a distance of $z$, the beam is detected by a receiver whose telescope has a circular aperture with radius $a_\mathrm{G}$. For the satellite's telescope, we assume a diameter of 10 cm (compatible with CubeSats). For the ground station, we take the telescope diameter from the Micius experiment (1 m) \cite{liao2017satellite}. To minimize the impact of turbulence, we only consider the downlink configuration (which can be compensated for using adaptive optics in the ground station) \cite{pirandola2021satellite}. In addition, configuring an uplink also entails addressing the challenge of having single-photon detectors on the satellite.\\
\indent Because of the inevitable free-space diffraction, the waist of the beam will expand during propagation. After travelling for a distance $z$, the beam has expanded to
\begin{equation}
\label{wG}
w_\mathrm{G}(z) = w_0 \sqrt{1+ \frac{z^2}{z_R^2}},
\end{equation}
where $z_R=\pi w_0^2 \lambda^{-1}$ is the Rayleigh range and $w_0=5$ cm is the beam radius at the exit of the satellite's telescope. The transmission (see Supplementary Section S1) through the receiver telescope's aperture $a_\mathrm{G}$ is given by
\begin{equation}
\eta_\mathrm{G}(z) := 1-\exp \left( -2\frac{a_\mathrm{G}^2}{w_\mathrm{G}^2}\right).
\end{equation}
In our calculations, we neglect the refraction effect of the atmosphere on the elongation of the beam path for different angles.
\subsection{Atmospheric extinction}           
The atmospheric extinction is caused by both aerosol absorption and Rayleigh/Mie scattering, which results in free-space propagation loss. To calculate the atmospheric transmission as a function of wavelength and zenith angle ($\theta_\mathrm{zen}$), we use an open-source code, libRadtran \cite{emde2016libradtran}, and obtain the atmospheric transmission $\eta_A$ for 5$^\circ$ steps in the range from -80$^\circ$ to +80$^\circ$ (see also Supplementary Section S2).
\subsection{Pointing and tracking losses}
The satellite is assumed to point its telescope toward the ground station and track it during the pass. This tracking is usually performed using a beacon laser from the ground station. Due to the involved distances, a point ahead angle needs to be considered to compensate for the finite speed of light. This point ahead angle is calculated for the satellite, based on its orbital data \cite{kaushal2017free} and then used as an offset to the angle of incidence of the beacon beam. For an orbit of 500 km altitude, this point ahead angle is up to 53 $\mu$rad. Both the beacon beam sensing as well as the point ahead mechanism are subject to errors. They are, however, assumed to be independent of each other and normally distributed with zero means.\\
\indent For these conditions, the link efficiency due to pointing and tracking errors, $\eta_\mathrm{pt}$, can be modeled as \cite{valentini2021analysis, chen1989impact} 
\begin{equation}
\eta_\mathrm{pt} = \exp(-G_\mathrm{tx} \theta_\mathrm{pt}^2),
\end{equation}
with the transmit antenna gain $G_\mathrm{tx} = (\frac{2 \pi a_\mathrm{S}}{\lambda})^2$ and the combined pointing and tracking error angle $\theta_\mathrm{pt}$ of the satellite. Note, $a_\mathrm{S} = 5$ cm is the clear transmitter aperture radius. For the calculation of the corresponding loss, a pointing and tracking error of 1 $\mu$rad is assumed \cite{liao2017satellite}. As the atmospheric beam spread affects the transmit antenna gain, we model this using the Hufnagel-Valley model 5/7 \cite{hufnagel1964modulation, valley1980isoplanatic} with a structure constant $A = 1.7 \times 10^{-14}$ m$^{-2/3}$ of the surface and the wind velocity $W = 21$ m/s. For the downlink scenario, however, the resulting beam spread is in the sub-$\mu$rad range and has therefore only a minor impact on the link efficiency. The details of the pointing and tracking error calculations are presented in Supplementary Section S3.
\subsection{Temporal filtering}
As the dark counts occur randomly, they can be suppressed by temporal filtering the detection events \cite{ko2018experimental}. This can be either done actively by gating the detectors or by post-selection. In the following, we consider a temporal width of the photons of $\tau=3$ ns (either governed by the excited state lifetime or the laser pulse length, depending on the light source). For a fair comparison, we are assuming that the filtering efficiency for a laser pulse and a single-photon emitter is identical (in practice they depend on the exact temporal distribution). For a single-photon emitter, the spontaneous emission process is described by an exponential distribution. The filtering efficiency for a gating time $\Delta t$ is given by the cumulative distribution function:
\begin{equation}
\eta_{\mathrm{temporal}} = 1-\exp\left(-\frac{\Delta t}{\tau}\right).
\end{equation}
For each parameter setting (i.e., wavelength and light source), the choice of $\Delta t$ is numerically optimized (see also Supplementary Section S4).
\subsection{Total transmission}  
The total transmission of the free-space communication link is given by the product of the specific transmission terms. To get the total detection efficiency $\eta$, this number needs to be multiplied by the efficiency of the single-photon detectors $\eta_{\mathrm{det}}$:
\begin{equation}
\eta = \eta_{\mathrm{det}} \eta_\mathrm{G} \eta_\mathrm{A} \eta_\mathrm{pt} \eta_{\mathrm{temporal}}.
\end{equation}
Note that we have neglected internal optical losses in the sender and receiver unit, which are typically small compared to the other link losses due to anti-reflection coatings and high-quality optics. For the wavelength-dependent efficiency of the single-photon detectors, we have extracted the specified efficiencies from the most commercially available single-photon detectors, including silicon and InGaAs/InP SPADs, as well as SNSPDs. As one is free in the detector choice, we have selected the most efficient detector for each wavelength (for details see Supplementary Section S5).
\subsection{Background noise}\label{subsec: background}     
The maximal communication distance of a discrete-variable QKD system is limited by the background noise that causes the typical sharp cutoff in the rate-distance dependency. This consists of the intrinsic detector dark count rate (DCR) and extrinsic noise sources such as the sun during daylight. The latter is the dominant term (at least during daylight) and depends on the operational setting (e.g., time of the day and direction of the link), as well as the receiver's aperture size $a_\mathrm{G}$ and field of view $\Omega_{fov}$ (collecting this background). Many QKD systems use temporal filtering \cite{ko2018experimental}, which reduces background noise as the detectors are only active for a time $\Delta t$ when a photon from the sender is expected (in practice this is often done by post-selection). The solar background is of course dependent on the wavelength $\lambda$ and spectral filter bandwidth $\Delta \lambda$. It is convenient to define the parameter
\begin{equation}
\Gamma_R =  \Delta t \Omega_{fov} a_\mathrm{G}^2.
\end{equation}
In our calculations, we assume $\Omega_{fov} = 10^{-10}$ sr \cite{pirandola2021satellite} and $a_\mathrm{G} = 50$ cm for the receiving telescope \cite{liao2017satellite}. Note that these parameters depend on the specific QKD system and are only exemplary here. Hence, the probability of background detection is given by
\begin{equation}
p_\mathrm{BG}(\lambda)= \eta_\mathrm{det}(\lambda) \Gamma_R H_\lambda ^\mathrm{sun}+\text{DCR}(\lambda)\Delta t.
\end{equation}
As already mentioned, the wavelength-dependent efficiencies $\eta_\mathrm{det}(\lambda)$ are extracted from commercially available single-photon detectors (see Supplementary Section S5). In addition, we take the detector dark count rate from the detector with the highest efficiency for each wavelength, making DCR also wavelength-dependent. $H_\lambda ^{sun}$ is the solar spectral irradiance integrated over the spectral filter bandwidth. Unless stated otherwise, we take $\Delta\lambda = 0.5$ nm for all wavelengths. As a filter system, a Fabry-P\'{e}rot etalon can be used, which can achieve polarization-independent filtering at normal incidence and the required linewidth while maintaining a sufficiently large free spectral range. The solar spectral irradiance is calculated using the libRadtran open-source code for a downlink when the sun is at the zenith angle of 45$^\circ$ and the solar azimuth is 0$^\circ$ (see Supplementary Section S6 for more details). We consider a fixed sun position as the flyover time (a few minutes) is small compared to the period of the sun (one day). The position of the sun (zenith angle of 45$^\circ$) represents some average daylight conditions between sunset/sunrise (best conditions) and noon (worst condition). The detector pointing follows the trajectory of the satellite and we calculate $H_\lambda ^{sun}$ also in steps of 5$^\circ$. During the night we omit the background light from stars and the moon and simply set $H_\lambda ^{sun}$ to zero.
\subsection{Photon detection and error}  
The following calculations of secret key rates follow the standard formalism in QKD \cite{Lo2005quantum}. When using a threshold detector (i.e., it can differentiate a vacuum from a non-vacuum state but not resolve the photon number), the detection probability of an $n$-photon Fock state given by
\begin{equation}
\eta_n = 1-(1-\eta)^n ~~~\forall ~n \in \mathbb{N}.
\end{equation}
If the sender (Alice) sends an $n$-photon state, then define $Y_n$ as the yield of that state, i.e., the conditional probability of a detection event occurring at the receiver (Bob) side. The term $Y_0$ denotes the yield from background events and is equal to $p_{BG}$. Considering that the background counts are independent of the detection of signal photons, $Y_n$ is provided by
\begin{equation}
Y_n = Y_0 + \eta_n - Y_0 \eta_n \approx Y_0 + \eta_n,
\end{equation}
where the last term is omitted in the approximation as both factors are typically small. The gain of the $n$-photon state $Q_n$ can be derived as
\begin{equation}
Q_n = Y_n p(n),
\end{equation}
for $p(n)$ being the probability of sending $n$ photons, i.e., the gain is the conditional probability of Alice sending an $n$-photon state that leads to a detection event for Bob. The error of the $n$-photon state, $e_n$, can be calculated as
\begin{equation}
e_n = \frac{e_0Y_0 + e_{int}\eta_n}{Y_n},
\end{equation}
and depends on the background term and the system intrinsic error $e_{int}$. The latter depends on the state preparation quality, channel depolarization, and optical alignment of the components. Here we take $e_{\mathrm{int}}=3$\%, which is a typical value for free-space QKD links \cite{vest2022quantum,liao2017satellite}. In addition, if the detectors have equal DCRs, then $e_0 = \frac{1}{2}$ (uncorrelated background). The total gain is described by
\begin{equation}
Q_\mu = \sum_{n=0}^{\infty} Y_n p(n),
\end{equation}
and the total quantum bit error ratio (QBER) $E_\mu$ can be calculated from
\begin{equation}
E_\mu Q_\mu = \sum_{n=0}^{\infty} e_n Y_n p(n).
\end{equation}
\subsection{Single-photon and weak coherent sources}
We now turn to calculating secret key rates as the defining performance metric for different light sources. In the simplest case, Alice can only send the signal states. The extractable secret bit per photon \cite{ma2005practical} for practical BB84-like protocols is 
\begin{multline}\label{eqn:extractablesecretKeyGeneral}
S \geq q \times \mathrm{max}\{-Q_{\mu}f(E_\mu)h_2(E_\mu)\\
+Q_1 (1-h_2(e_1)), 0\},
\end{multline}
where $h_2(p)$ is the binary Shannon entropy
\begin{equation}
h_2(p)=-p \log_2(p)- (1-p) \log_2(1-p).
\end{equation}
For symmetric basis encoding (standard BB84) we use the sifting ratio $q=\frac{1}{2}$ and a constant error correction efficiency of $f(E_\mu)= 1.22$ from the GYS experiment \cite{gobby2004quantum}. In the weak version \cite{wang2005beating} of the GLLP method \cite{gottesman2004security}, the extractable secret information is
\begin{multline}\label{eqn:extractablesecretKey}
S \geq q \times Q_{\mu} \times \mathrm{max}\{-f(E_\mu)h_2(E_\mu)\\
+\Omega(1-h_2(\frac{E_\mu}{\Omega})), 0\}.
\end{multline}
The parameter $\Omega$ is the fraction of `untagged' photons (i.e., the fraction of Bob's detection events originating from single-photons sent by Alice) \cite{Lo2005quantum}. This is given by
\begin{equation}
\Omega = 1-p(n>1)/ Q_\mu,
\end{equation}
where $p(n>1)$ is the multi-photon probability. This is a worst-case assumption, where one assumes that all multi-photon states sent by Alice are received by Bob.\\
\indent We can use the formulas above to calculate the secret key rates for light sources which can only emit signal states, such as single-photon sources or weak coherent state sources (lasers). For the latter, the number of photons in a laser pulse is Poisson distributed with mean photon number $\mu$, i.e.,
\begin{equation}
p(n)=\frac{\mu^n}{n!}e^{-\mu}.
\end{equation}
An ideal single-photon source has $p(1)=1$. The non-ideal SPS is parametrized by its mean photon number $\overline{n}$ and second-order correlation function at zero time delay $g^{(2)}(0)$. For the multi-photon probability, an upper bound of
\begin{equation}
p(n>1)\leq\frac{1}{2}\overline{n}^2g^{(2)}(0)
\end{equation}
can be found \cite{PhysRevA.66.042315}. For the realistic single-photon source (`our SPS') we take the parameters from a previous experiment \cite{vogl2019compact, PhysRevResearch.3.013296} and extract the following parameters:
\begin{equation*}
\begin{gathered}
p(0)=1-p(1)-p(n>1)=4.861\times10^{-1},\\
p(1)=5.13\times10^{-1},\\
p(n>1)=\frac{1}{2}\overline{n}^2g^{(2)}(0)=8.48\times10^{-4},\\\
\overline{n}= p(1) +  \sum_{n=2}^{\infty} n p(n)=5.147\times10^{-1},\\\
g^{(2)}(0)=6.4\times10^{-3}.\
\end{gathered}
\end{equation*}
\subsection{Two-decoy states source}
The basic BB84 protocol gains its security that non-orthogonal states are not perfectly distinguishable. The photon number states are also non-orthogonal, hence, an eavesdropper cannot distinguish two different coherent states with a randomized phase. The yield for Fock states then becomes independent of the intensity. This is exploited in QKD with decoy states, as it allows for the detection of photon number splitting attacks \cite{hwang2003quantum, Lo2005quantum}. These decoy protocols are the most efficient laser-based QKD protocols available to date and have therefore been implemented in satellite-based QKD \cite{liao2017satellite}.\\
\indent In practice, it is sufficient to use two decoy states to find better bounds for $Y_1$ and $e_1$ (i.e., a better bound that was found in the previous section) \cite{ma2005practical}. Due to simplicity, one can choose the decoy intensities (mean photon number) $\nu_1$ and $\nu_2=0$ (i.e., vacuum). In this case, Eq.\ \ref{eqn:extractablesecretKeyGeneral} can be rewritten as
\begin{multline}\label{eqn:extractablesecretKeyTDS}
S \geq q \times \mathrm{max}\{- Q_{\mu}f(E_\mu)h_2(E_\mu)\\
+Q_1^{\mathrm{L}, \nu_1, \nu_2} (1-h_2(e_1^{\mathrm{U}, \nu_1, \nu_2})), 0\}
\end{multline}
where L and U denote the lower and upper bounds, respectively \cite{ma2005practical}. In the limits of a low background rate ($Y_0 \ll \eta$) and a small transmittance ($\eta \ll 1$), this can be simplified to
\begin{equation}
S \approx - \eta \mu f(e_{int})h_2(e_{int})+\eta \mu e^{-\mu} (1-h_2(e_{int}))
\end{equation}
This quantity is optimized for an an optimal $\mu = \mu_{\mathrm{opt}}$ which satisfies
\begin{equation}
(1- \mu) \mathrm{exp}(-\mu) = \frac{f(e_{int})h_2(e_{int})}{1-h_2(e_{int})}
\end{equation}
As the sender is free in the choice of $\mu$, this has to be optimized for every link parameter set.
\section{Finite secret key length analysis}
We now extend the previous asymptotic analysis to the finite key regime to evaluate a more rigorous QKD performance of the TDS and our SPS protocols previously shown. The widely applied asymptotic analysis is a good approximation under a sufficiently large number of received signals. However, in practice, the size of these samples is constrained by the satellite overpass transmission time due to finite-block size effects. Thus, the secret key that is extracted from short blocks of received signals deviates from the asymptotic expected values. Therefore, the statistical fluctuations of the estimated parameters cannot be neglected. They must be bounded, thereby introducing a correction term. The implemented bounds in this work, namely the multiplicative Chernoff bound and the random sampling without replacement bound, have been proven to be tighter bounds than the other mathematical bounds \cite{yin2020tight}.
\subsection{Optimization of the protocols}
\begin{table}[bt]
\caption{Fixed QKD system parameters. The intrinsic error and the source repetition rate are set conservatively compared to certain QKD demonstrations.}
\label{tab:SysPar}
\centering
\begin{ruledtabular}
\begin{tabular}{lll}
\textbf{Description} & \textbf{Parameter} & \textbf{Value} \\
\hline
Intrinsic error & $e_{\mathrm{int}}$ & 0.03 \\
Source repetition rate & $R_{\mathrm{rate}}$ & $10^{8}$ Hz \\
Detection time window & $\Delta t$ & 1 ns \\
Secrecy parameter & $\varepsilon_{\mathrm{sec}}$ & $10^{-9}$ \\
Correctness parameter & $\varepsilon_{\mathrm{cor}}$ & $10^{-15}$ 
\end{tabular}
\end{ruledtabular}
\end{table}
In previous sections, the standard BB84 was used, where none of the protocol parameters is optimized, equal basis bias is assumed and the key is extracted from the signal states with intensity $\mu$. In this section, to maximize the attainable secret key the efficient BB84 protocol is implemented where the $X$ basis is exclusively used for the key generation while the $Z$ basis is used for the phase error rate estimation. Thus, in the TDS the highest key length is generated by optimizing the following parameters, the basis bias $p_{\mathrm{X}}$, the intensities of the two sources $\mu$ and $\nu_1$, the probabilities of choosing each source $p_{\mu}$ and $p_{\nu_1}$, and the transmission time window $\delta t$. Notice that we set the other decoy state intensity as vacuum $\nu_2 = 0$. Please note that, here only WCS is assumed to be used for two-decoy case. For the SPS, there are fewer protocol parameters, hence, we generate an optimized secret key length (SKL) by finding the optimal $p_{\mathrm{X}}$, $\delta t$ and the pre-attenuation of Alice's source $\eta_{\mathrm{att}}$. The latter parameter quadratically reduces the multi-photon emission probability $p \left( n>1 \right)$ of the SPS, while linearly decreasing the click probability $p_{\mathrm{click}}$ and increasing the maximum tolerable losses. In both cases, TDS and SPS, the number of sent pulses $n_{\mathrm{S}}$ is determined by the source repetition rate $R_{\mathrm{rate}} = 100$ MHz. To perform a fair comparison, we consider a detection time window $\Delta t = 1$ ns for all wavelengths and the same fixed parameter for both QKD protocols, see Table \ref{tab:SysPar}. To simulate satellite-to-ground QKD communication and optimize the parameters, we use the SatQuMa simulation toolkit \cite{sidhu2021satellite}. While TDS was already implemented in it, we introduced the SPS protocol.\\
\indent The optimization works as follows: first, the satellite altitude $h$ and the maximum elevation angle of the overpass $\theta_{\mathrm{max}}$ are fixed. Thus, we obtain a fixed total transmission time for the single overpass and an associated link efficiency for each time slot. Then, the optimization of the SKL over the protocol parameters, $\{p_{\mathrm{X}},\mu,\nu_1,p_{\mu},p_{\nu}\}$ for TDS and $\{p_{\mathrm{X}},\eta_{\mathrm{att}}\}$ for SPS, is run for each possible time windows, from $-\delta t$ to $+\delta t$ (the time of the whole overpass is the maximum window), to find the one resulting in the highest SKL. This $\delta t$ optimization is done because when the QBER is extremely high at lower elevation angles taking a shorter finite block than the whole pass gives us greater SKL.
\subsection{Two-decoy state source}
In this section, we highlight the main aspects of the TDS following the methods in \cite{sidhu2022finite} and utilizing the derived yields in \cite{lim2014concise}. For TDS, to model the statistical uncertainties of the estimated parameters, we introduce correction terms that depends on the collected finite statistic. Thus, for TDS we bound the number of events in either basis or intensity $n_{\mathrm{X(Z),}\mu(\nu_1)}$, i.e., the gain and the number of errors $m_{\mathrm{X(Z),}\mu(\nu_1)}$ for each basis or intensity, using the Chernoff bound as follows
\begin{align}
    n_{\mathrm{X(Z),}\mu(\nu_1)}^{\pm} & = e^{\mu(\nu_1)} \left[ n_{\mathrm{X(Z),}\mu(\nu_1)} \pm \delta_{n_{\mathrm{X(Z),}\mu(\nu_1)}}^{\pm} \right] / p_{\mu(\nu_1)} \\
    m_{\mathrm{X(Z),}\mu(\nu_1)}^{\pm} & = e^{\mu(\nu_1)} \left[ m_{\mathrm{X(Z),}\mu(\nu_1)} \pm \delta_{m_{\mathrm{X(Z),}\mu(\nu_1)}}^{\pm} \right] / p_{\mu(\nu_1)} 
\end{align}
where the deviation $\delta^{\pm}$ depends on the number of events or errors and on the secrecy parameter $\varepsilon_{\mathrm{sec}}$,
\begin{equation}
    \delta^{+}_{x} = \beta + \sqrt{2 \beta x + \beta^{2}} , \hspace{0.3cm} \delta^{-}_{x} = \frac{\beta}{2} + \sqrt{2 \beta x + \frac{\beta^2}{4}},
\end{equation}
where $\beta = -\log_{\mathrm{e}} \left(\varepsilon_{\mathrm{PE}}\right)$ with parameter estimation $\varepsilon_{\mathrm{PE}} = \varepsilon_{\mathrm{sec}} / 21$ being $x$ either $n_{\mathrm{X(Z),}\mu(\nu_1)}$ or $m_{\mathrm{X(Z),}\mu(\nu_1)}$. Notice that the upper and lower bounds of these quantities are chosen to lower bound the vacuum $s^{\mathrm{L}}_{\mathrm{X,0}}$ and single-photon $s^{\mathrm{L}}_{\mathrm{X,1}}$ yields. In the phase error rate estimation, Bob randomly selects a subset of the received signals, that may not be fully representative of the entire transmission, and he measures each qubit once. Here is where the random sampling without replacement problem arises and a correction term $\gamma$ is introduced to compensate for any potential statistical deviation due to the subset selection. Thus, we estimate the upper bound of the phase error rate in the X basis using the Z basis as follows 
\begin{equation}
    \phi^U_{\mathrm{X}} = \dfrac{v^U_{\mathrm{Z,1}}}{s^L_{\mathrm{Z,1}}} + \gamma \left(s^L_{\mathrm{X,1}}, s^L_{\mathrm{Z,1}}, \dfrac{v^U_{\mathrm{Z,1}}}{s^L_{\mathrm{Z,1}}}, \dfrac{\varepsilon_{sec}}{21} \right) ,
\end{equation}
where the $v^U_{\mathrm{Z,1}}$ is the upper bound of the single-photon errors.

The finite SKL formula for the BB84 TDS is given by
\begin{multline}
    \ell_{2D} = s^{\mathrm{L}}_{\mathrm{X,0}} + s^{\mathrm{L}}_{\mathrm{X,1}} \left[ 1 - h_2 \left(\phi^{\mathrm{U}}_{\mathrm{X}} \right) \right]\\
    - \lambda_{\mathrm{EC}} - 6 \log_{\mathrm{2}} \dfrac{21}{\varepsilon_{\mathrm{sec}}} - \log_{\mathrm{2}} \dfrac{2}{\varepsilon_{\mathrm{cor}}} 
\end{multline}
where $\lambda_{\mathrm{EC}} = 1.22 \, n_{\mathrm{X}} \, h_{\mathrm{2}}(E_{\mathrm{X}}) $ is the number of bits used in the error correction process. To generate the key, we now use the events in the X basis such as $n_{\mathrm{X}}$ and the QBER $E_{\mathrm{X}}$, instead of the events caused by the intensity $\mu$ as in standard BB84. The remaining terms represent the bits used to verify the correctness and secrecy of the protocol, ensuring that the protocol is $\varepsilon$-secret (where $\varepsilon = \varepsilon_{\mathrm{sec}} + \varepsilon_{cor}$) within the universally composable security framework.
\subsection{Single-photon source}
In this section, we follow the non-decoy method for a SPS described in \cite{morrison2023single}. As only one source is used, we simply calculate the non-multiphoton yield (lumping together the vacuum and single-photon yield) on either basis as
\begin{equation}
    n_{\mathrm{X(Z),nmp}} = n_{\mathrm{X(Z)}} - n_{\mathrm{X(Z),mp}}
\end{equation}
where $n_{\mathrm{X(Z)}}$ is the total number of received signals and $n_{\mathrm{X(Z),mp}} = n_{\mathrm{S}} \, p_{\mathrm{X}} \, p \left( n>1 \right)$ is the sifted multiphoton emission. The number $n_{\mathrm{X(Z)}}$ is directly observed by Bob, so no bound is needed. However, the true number of multiphoton emissions when both parties match their basis $n_{\mathrm{X(Z),mp}}$ cannot be observed. Therefore, we apply the Chernoff bound's variant used in TDS to upper bound the estimated $n_{\mathrm{X(Z), mp}}$ with error $\varepsilon_{\mathrm{PE}}$ as
\begin{equation}
    n^{\mathrm{U}}_{\mathrm{X(Z),mp}} = \left(1 + \delta^{\mathrm{U}}  \right) n_{\mathrm{X(Z),mp}} ,
\end{equation}
where $\delta^{\mathrm{U}} = \left( \beta + \sqrt{8 \beta n_{\mathrm{X(Z),mp}} + \beta^2}\right) / (2 n_{\mathrm{X(Z),mp}})$ is the correction term and we have $\beta = -\log_{\mathrm{e}} \left( \varepsilon_{\mathrm{PE}} \right)$ with $\varepsilon_{\mathrm{PE}} = 2 \, \varepsilon_{\mathrm{sec}} / 3$. Thus, we directly get the lower bound on the non-multiphoton yield $n^{\mathrm{L}}_{\mathrm{X(Z), nmp}} = n_{\mathrm{X(Z)}} - n^{\mathrm{U}}_{\mathrm{X(Z), mp}}$. The finite SKL for the general BB84 protocol reads as
\begin{multline}
    \ell_{\mathrm{ND}} = n^{\mathrm{L}}_{\mathrm{X,nmp}} \left[1-h_{\mathrm{2}}\left(\phi^{\mathrm{U}}_{\mathrm{X}}\right)\right]\\
 -\lambda_{\mathrm{EC}} - 2\log_{\mathrm{2}} \frac{3}{\varepsilon_{\mathrm{sec}}} - \log_{\mathrm{2}} \frac{2}{\varepsilon_{\mathrm{cor}}}.
\end{multline}
Here, $\lambda_{\mathrm{EC}}$ and the upper bound of the phase error rate are derived as in TDS,
\begin{equation}
    \phi^U_{\mathrm{X}} = \dfrac{m_{\mathrm{Z}}}{n^L_{\mathrm{Z,nmp}}} + \gamma \left(n_{\mathrm{X}}, n_{\mathrm{Z}}, \dfrac{m_{\mathrm{Z}}}{n^L_{\mathrm{Z,nmp}}}, \dfrac{4 \, \varepsilon_{\mathrm{sec}}}{6} \right) ,
\end{equation}
where we now consider the total number of errors in the Z basis $m_{\mathrm{Z}}$ for the phase error. The secrecy term comes from The \textit{Leftover Hashing Lemma} related to the privacy amplification process when Alice and Bob apply a hash function with a failure probability of $\varepsilon_{\mathrm{PA}} = \varepsilon_{\mathrm{sec}} / 6$ and the other logarithmic term corresponds to the error correction process.
\subsection{Annual secret key length}
The expected annual SKL is defined as the generated key during a whole earth's rotation around the sun equivalent to $t_{\mathrm{ER}} \approx 31558140$ s and here we use the estimation method of \cite{sidhu2022finite}. For a given satellite altitude (that we set at 500 km) we estimate the maximum SKL for each non-zenith overpass (maximum elevation angle $\theta^{\mathrm{max}}_{\mathrm{elev}} < 90^{\circ}$). The maximum elevation angle of the pass can be expressed as the ground track off-set $d$, i.e., the distance from the ground station to where the satellite ground track crosses the longitudinal circumference that goes through the station. Thus, we represent the SKL vs.\ $d$ for single passes, and we integrate the area below each curve as
\begin{equation}\label{eqn:SKL_int}
    \mathrm{SKL}^{\mathrm{Day(Night)}}_{\mathrm{int}} = 2 \int^{d_{\mathrm{max}}}_{0} \mathrm{SKL}^{\mathrm{Day(Night)}}_{d} \,\, \mathrm{d}d
\end{equation}
where $\mathrm{SKL}^{\mathrm{Day(Night)}}_{d}$ is the SKL per pass as a function of the ground tack off-set, and $d_{\mathrm{max}}$ is the maximum tolerable ground track off-set. This integral represents the area below the secret key curve in bits-meter (bm) units. The factor 2 comes from the fact that we have negative values for the ground track off-set that generates the same key as the positive $d$. Thus, since we do not model the orbits over time, we estimate the average SKL of a single orbit by dividing this area by the line of longitude of the station latitude $L_{\mathrm{lat}}$.\\
\indent Assuming a satellite in a sun-synchronous orbit and neglecting weather, we estimate the annual secret key using blocks formed by single passes during either daylight or night as
\begin{equation}\label{eqn:SKLannualDayNight}
    \overline{\mathrm{SKL}}^{\, \mathrm{Day(Night)}}_{\, \mathrm{annual}} = \dfrac{N^{\mathrm{annual}}_{\mathrm{orbits}}}{L_{\mathrm{lat}}} \, \mathrm{SKL}^{\mathrm{Day(Night)}}_{\mathrm{int}} .
\end{equation}
If the ground station latitude is close to the poles the previous approximation to estimate the annual SKL is not valid \cite{mazzarella2020quarc}. In this work, we consider the location of the station at the equator. This corresponds to the longitudinal circumference of the station latitude $L_{\mathrm{lat}} \approx 4.008 \times 10^{7} $ m, modeling the Earth as a world geodetic system (WGS) 84 ellipsoid. Lastly, the total number of orbits during the earth's rotation around the sun is $N^{\mathrm{annual}}_{\mathrm{orbits}} = t_{\mathrm{ER}} /  T_{\mathrm{S}} \approx 5567$ \footnote{For simplicity in our notation we use the word annual, we refer to the time earth takes to do a whole rotation around the sun, not to the 365 days approximation}. In one orbit, the satellite crosses the line of latitude of the ground station twice, once during daylight and once during the night.  Therefore, to calculate the total annual secret key we add together the SKL during daylight and night
\begin{equation}\label{eqn:SKLannual}
    \overline{\mathrm{SKL}}_{\, \mathrm{annual}} = \overline{\mathrm{SKL}}^{\, \mathrm{Day}}_{\, \mathrm{annual}} + \overline{\mathrm{SKL}}^{\, \mathrm{Night}}_{\, \mathrm{annual}}.
\end{equation}
\section{\label{sec:results}Results}
 To mitigate time-consuming computational complexity and consider the wide wavelength range we cover ($400-1700$ nm), we implement the standard BB84 protocol in the asymptotic regime first, where there is no optimization and uncertainties involved, to find the optimal operating wavelength that could yield the maximum key rate enhancement for our SPS or the TDS protocol. With this simplistic model established, we can calculate the secret key rate spectrum using Eqs.\ \ref{eqn:extractablesecretKey} and \ref{eqn:extractablesecretKeyTDS}, thereby identifying the most efficient wavelength for daylight QKD. We also calculate this spectrum for night conditions to obtain the effective average key rate for a full satellite orbit. As previously noted, we integrate over an entire satellite pass, necessary as the link efficiency varies during the flyover, e.g., when the satellite crosses the horizon, the slant distance increases and the path through the atmosphere's turbulent layers extend.
     \begin{figure*}
        \centering
        \includegraphics[width=14cm]{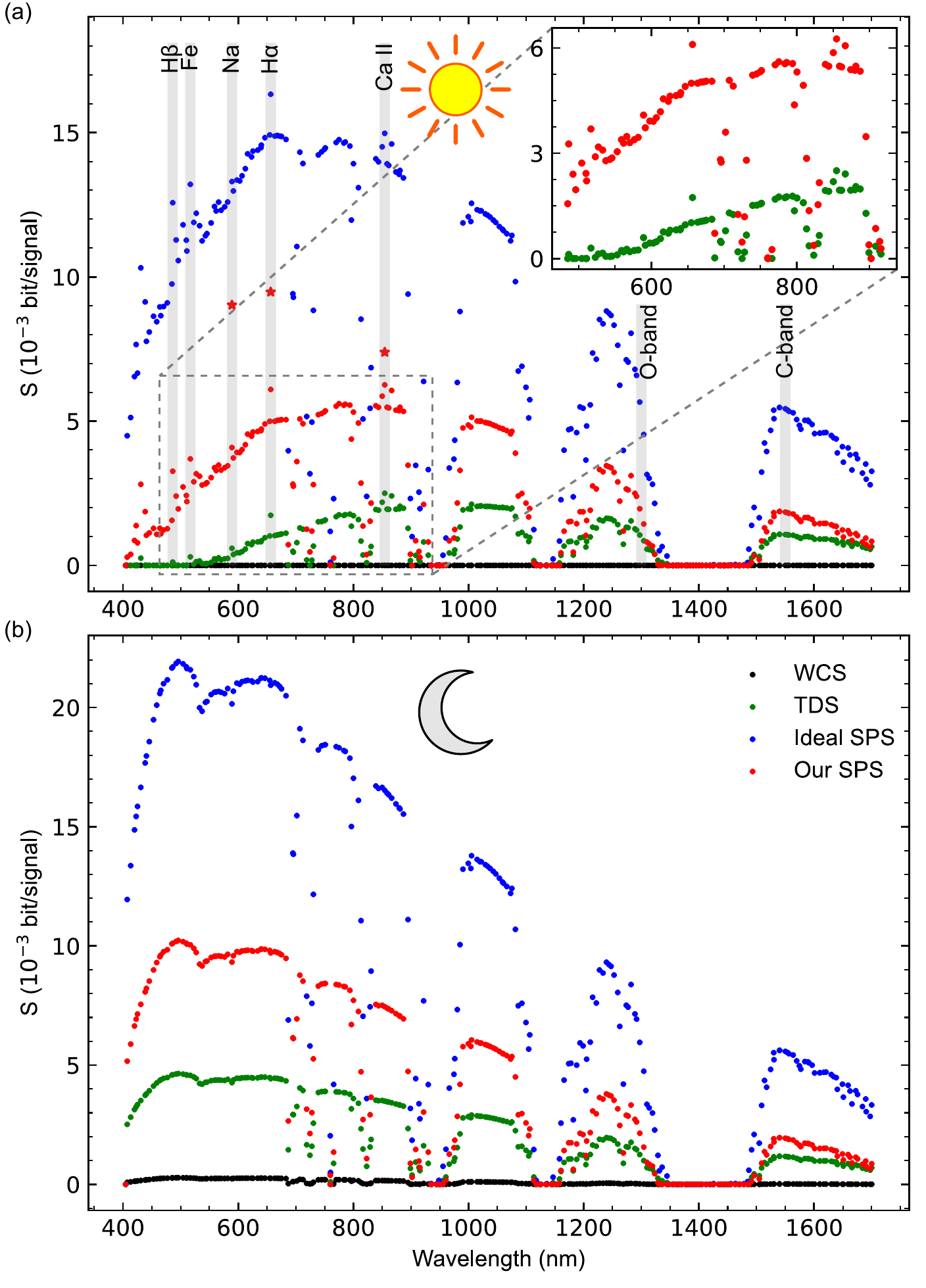}
        \caption{The extractable secret bit per signal for one pass of a satellite with an altitude of 500 km (a) in daylight and (b) at night. During daylight, some Fraunhofer lines (labeled with the responsible elements) stand out due to their reduced background. The `$\star$' symbols mark the key rate if special Fraunhofer filters (commercially available for selected lines) typically used in solar astronomy are implemented. The detection time was set to 6 ns as it was the optimized time for H\textalpha\ line. During the night (with no background) there is no specific advantage at special wavelengths, however, the visible wavelengths perform better than infrared ones. The asymptotic model is employed here.}
        \label{fig:K_lda_6ns}
    \end{figure*}
\indent The results in the spectral range from 400 up to 1700 nm are presented in Fig.\ \ref{fig:K_lda_6ns}(a) for a satellite altitude of 500 km during daylight. Certain Fraunhofer lines, labelled by the responsible elements (hydrogen, calcium, sodium, and iron), show distinct peaks for our performance metric (extractable secret bits per signal). For the sake of clarity, not all lines are labelled (considering the existence of several thousand Fraunhofer lines). Later, when considering the finite-key effect, we will use a 1 ns detection time, as it is more favourable given the higher operational repetition rate. As outlined in Supplementary Section S5, for a 1 ns detection time, the H\textalpha\ line outperforms the Ca~II line for our SPS, while for the TDS, the Ca~II line shows slightly better performance. That is why we tried to maximize the extractable secret bits per signal for H\textalpha\ line by varying the detection time in our asymptotic model. Therefore, in Fig.\ \ref{fig:K_lda_6ns} we assume a 6 ns detection time for the calculation.  For an ideal SPS, the H\textalpha\ ($\lambda= 656.448$ nm) line is the most efficient, whereas for our realistic SPS and the TDS, the Ca~II ($\lambda= 854.445$ nm) line shows higher efficiency given the 6 ns detection time assumed in our asymptotic model (see Supplementary section S5 to observe the detection time dependency of the key rate). Furthermore, our realistic SPS outperforms the TDS in terms of the number of available wavelengths yielding a key rate and in terms of an overall higher key rate across all wavelengths. Considering a fixed 6 ns detection time, our SPS achieves a 3.5 (2.5) times higher secret key rate compared to the TDS for the H\textalpha\ (Ca~II) line. Intriguingly, operating a TDS-based QKD system at the H\textalpha\ line instead of the C-band offers 1.6 times higher extractable secret bits per signal. These observations underline the importance of selecting the operating wavelength and the source type.\\
\indent Notably, for both source types the telecom O- and C-band perform worse compared to lower wavelengths. This is in contrast to previous proposals using telecom wavelengths for daylight QKD \cite{liao2017long, gong2018free, avesani2021full}. We will go into more detail about the QKD performance of telecom C-band in the following paragraphs. Using a WCS in contrast, QKD is impossible during daylight (for our chosen conditions). So far, we have assumed a constant spectral filter bandwidth of 0.5 nm. There are, however, commercial Fraunhofer line filters available that are used in solar observation. These filters have a typical linewidth of 0.05 nm and can increase the secret key rate quite dramatically through a higher suppression. For our SPS, the secret key rates for these cases are marked with `$\star$' symbols in Fig.\ \ref{fig:K_lda_6ns}(a). It is worth noting that the performance increase is much more prominent for the H\textalpha\ line compared to the Ca~II line. This is due to the fact that at 656 nm there is intrinsically more sun light or in other words, the tighter filters can suppress more background in case of the H\textalpha\ line. However, as this is biasing the results toward the Fraunhofer lines used in astronomy, we are not considering this option any further. For sources besides our SPS, refer to Fig. S9 to see the results of the secret key rates when assuming filters with a 0.05 nm linewidth.\\
    \begin{figure*}[t]
        \includegraphics[width=17cm]{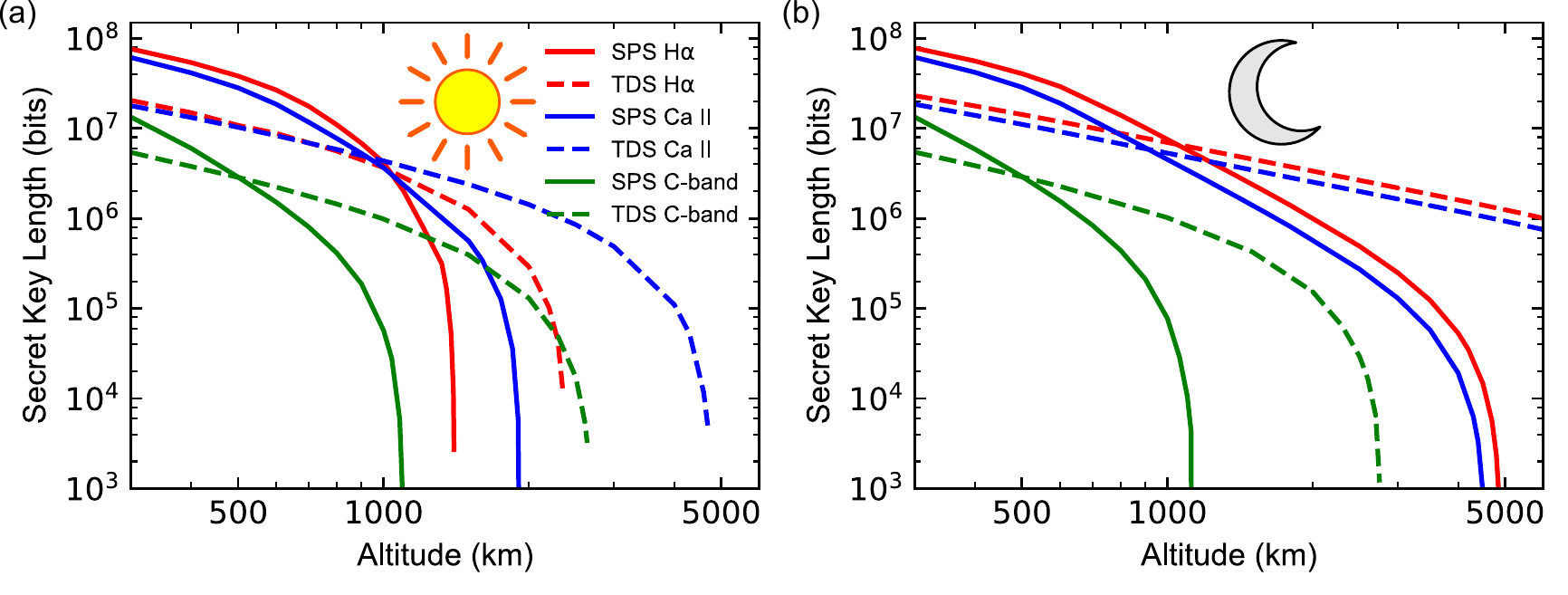}
        \caption{The secret Key Length (bits) as a function of the satellite altitude considering a zenith-pass for our single-photon source (SPS) and two-decoy source (TDS) emitting at H\textalpha\ (656.448 nm), Ca~II (854.445 nm), and C-band (1550.027 nm) (a) in daylight and (b) at night. Finite secret key length analysis is employed here.}
        \label{fig:SKL_altitude}
    \end{figure*}
    \begin{figure*}[t]
        \includegraphics[width=17cm]{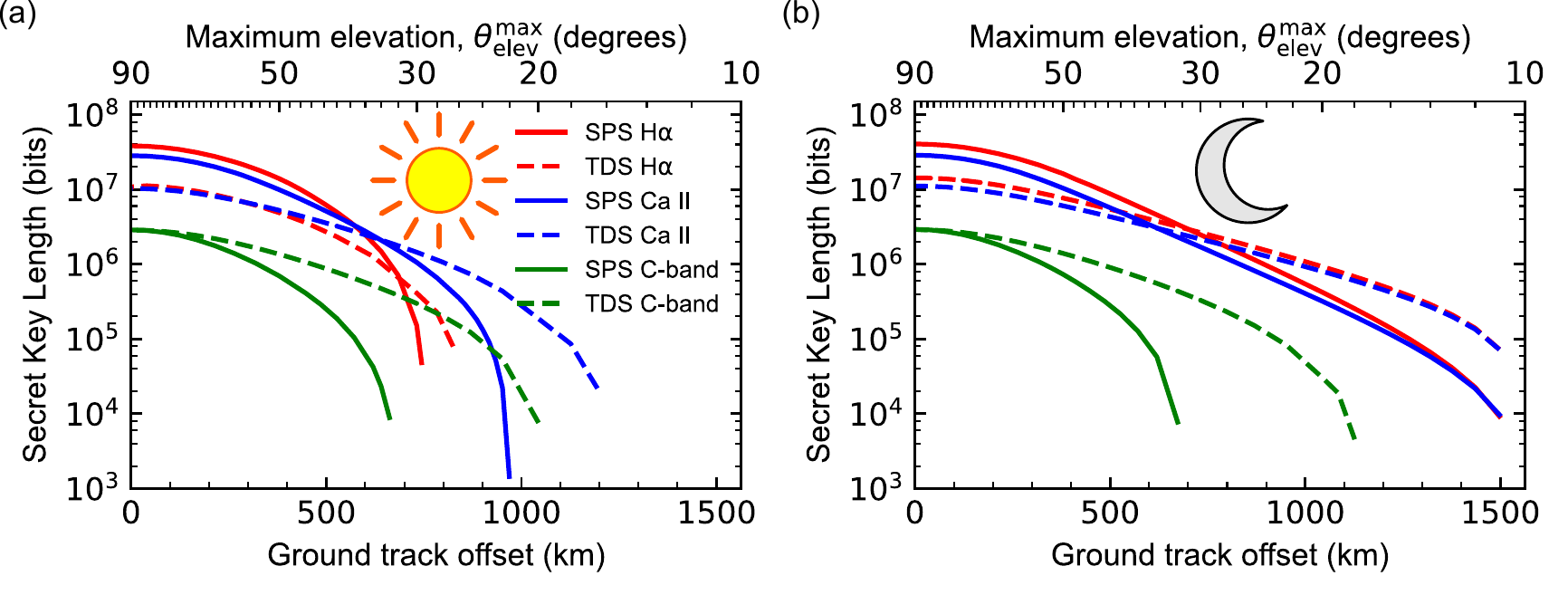}
        \caption{The secret Key Length (bits) as a function of the ground track offset (maximum elevation, $\theta_\mathrm{elev}^\mathrm{max}$) for our single-photon source (SPS) and two-decoy source (TDS) emitting at H\textalpha\ (656.448 nm), Ca~II (854.445 nm), and C-band (1550.027 nm) (a) in daylight and (b) at night. Finite secret key length analysis is employed here.}
        \label{fig:SKL_track_offset}
    \end{figure*}
\indent Obviously, using the Fraunhofer lines at night offers no advantage, as is evident by a `smoother' spectrum in Fig.\ \ref{fig:K_lda_6ns}(b). Additionally, overall secret key rates are higher during the night due to negligible background and for all sources (including WCS) the visible and infrared wavelengths become more efficient.\\
\indent As we now have identified the ideal wavelength (H\textalpha\ line) for 1 ns of detection time, we are now modelling the link in more detail. Note that we show the same asymptotic analysis also for the Ca~II line and the telecom C-band in Supplementary Section S7. The distance between a satellite orbiting at a 500 km altitude and the ground station is shown in Fig.\ S7. It takes about 420 s for the satellite to go from the zenith angle -80$^\circ$ to +80$^\circ$, while the slant distance varies from 1700 km to a minimum of 500 km, as we assumed a zenith-passing orbit. The resulting secret key rates and associated QBERs during the pass (for our source) are shown in Fig.\ S8(a) and (b), respectively. During the night, the secret key rates are higher, which is due to the lower QBER. The time window of a non-zero quantum link is then limited by the transmission through the atmosphere and the key rate drops to zero for long slant distances. The QBER is still nearly constant during the complete contact time. Interestingly, the QBER during daylight reaches the theoretical Shannon limit (11\%), which is the maximally tolerable QBER (with an ideal apparatus) to achieve a non-zero secret key rate. This reduces the active link time and therefore the overall efficiency of the pass compared to the nighttime case.\\
\indent The question arises, which altitudes would be possible for such a scenario (for $\lambda= 656.448$ nm). We model the extractable secret bit per signal for a satellite at a fixed zenith position ($\theta_{\mathrm{zen}}=0$, i.e., peak key rate) as a function of its altitude for both daylight and at night (see Fig.\ S8 (c) and (d), respectively). Here we again provide the results for all source types. We can conclude that our SPS outperforms WCS and TDS in daylight for all distances. The maximum altitude for which secret bits can be exchanged is beyond 1000 km. At night, our SPS achieves better results compared to other sources for up to 1700 km and the maximum altitude to have an operational quantum channel is beyond 2000 km. For very long distances, however, TDS becomes more efficient during the night. This has been reported before in simulations \cite{vogl2019compact} and is because decoy protocols are more loss-tolerant when the background noise is low. As expected, the ideal SPS outperforms all other light sources. For the results of the altitude calculations for the Ca~II line and telecom C-band see Supplementary Section S7.\\
    \begin{table*}[t]
        \caption{Expected annual secret key length with the ground station at the equator, i.e.\ $\scriptstyle \overline{\mathrm{SKL}}^{\, \mathrm{equator}}_{\, \mathrm{annual}} = 13.89 \times 10^{-4} \left( \overline{\mathrm{SKL}}^{\, \mathrm{Day}}_{\, \mathrm{annual}} + \overline{\mathrm{SKL}}^{\, \mathrm{Night}}_{\, \mathrm{annual}} \right)$ for each source during daylight and night time operation.}
        \label{tab:Annual_SKL}
        \begin{ruledtabular}
        \begin{tabular}{ccccc}
        \textbf{Source}  & \textbf{$\eta^{\mathrm{sys}}_{\mathrm{loss}}$ (dB)} & \textbf{$\overline{\mathrm{SKL}}^{\, \mathrm{Day}}_{\, \mathrm{annual}}$ (Gb)} & \textbf{$\overline{\mathrm{SKL}}^{\, \mathrm{Night}}_{\, \mathrm{annual}}$ (Gb)} & \textbf{$\overline{\mathrm{SKL}}^{\, \mathrm{equator}}_{\, \mathrm{annual}}$ (Gb)} \\
        \hline
        SPS - H\textalpha\ & 17.55 & 3.4829 & 4.0934 & 7.5763 \\
        TDS - H\textalpha\ &  17.55 & 1.1411 & 1.8912 & 3.0323 \\
        SPS - Ca~II &  18.60 & 2.6308 & 2.7958 & 5.4266 \\
        TDS - Ca~II&  18.60 & 1.2383 & 1.4972 & 2.7355 \\
        SPS - C-band &  23.31 & 0.2215 & 0.2278 & 0.4493 \\
        TDS - C-band &  23.31 & 0.3204 & 0.3308 & 0.6512 \\
        \end{tabular}
        \end{ruledtabular}
    \end{table*}
\indent Previously, we have identified two suitable Fraunhofer lines, H\textalpha\ and Ca~II, using standard BB84 in the asymptotic limit and we have compared their key generation with the commonly used telecom C-band for different detection times. In order to make more realistic predictions regarding the satellite-based QKD performance of these candidates, we have examined the impact of finite key effects. We have also implemented the optimized efficient-BB84 instead of the standard BB84 to compare the maximum attainable key by each candidate, see Section III A. Fig.\ \ref{fig:SKL_altitude} illustrates SKL as a function of satellite altitude for a zenith-pass, with our SPS and a TDS emitting at H\textalpha, Ca~II, and C-band wavelengths, both during the day and at night. When considering daytime operations, it is evident that H\textalpha\ and Ca~II Fraunhofer lines supersede the telecom C-band in performance for the altitudes smaller than 1240 km. Notably, our SPS surpasses the TDS up to an altitude of 1000 km for both H\textalpha\ and Ca~II lines. If we consider the C-band, our SPS performs better than a TDS up to 500 km altitude. This pattern is also observed during nighttime operations. However, at overly high altitudes, the TDS surpasses our SPS for both day and night. Moreover, in daylight, there is an advantage of the TDS C-band for altitudes higher than 1240 km for SPS H\textalpha, 1500 km for SPS Ca~II, and 2000 km for TDS H\textalpha. The improvement of the C-band in these regions occurs because it has the lowest background probability, e.g., $p_\mathrm{BG}^{\mathrm{C-band}} \approx 3.52 \times 10^{-6}$ at zenith, amongst the three wavelengths for 1 ns of detection time, e.g., $p_\mathrm{BG}^{\mathrm{H\alpha}} \approx 1.77 \times 10^{-4}$ and $p_\mathrm{BG}^{\mathrm{Ca \, II}} \approx 4.29 \times 10^{-5}$ at zenith, see also Supplementary Fig.\ S6. This together with the TDS make C-band tolerate higher altitudes in daylight communication.\\
\indent Most satellite overpasses will not cross over the zenith of the ground station. To extract the maximum amount of secret key for each overpass, the QKD system needs to operate even at low maximum elevation angles $\theta_\mathrm{elev}^\mathrm{max}$, although we impose a minimum elevation angle $\theta_\mathrm{elev}^\mathrm{min} = 10^{\circ}$ (ground track offset $d_\mathrm{min} \approx 1563$ km) for transmission. To evaluate all possible non-zenith passes, we represent the SKL vs the ground track offset and the maximum elevation angle of the overpass, see Fig.\ \ref{fig:SKL_track_offset}. As expected, when $\theta_\mathrm{elev}^\mathrm{max}$ decreases the extracted key is reduced. During the night, the C-band does not show any improvement over our SPS. During daylight operation, our SPS emitting at H\textalpha\ line outperforms all other alternatives up to a 600 km ground track offset ($\theta_\mathrm{elev}^\mathrm{max} > 40^{\circ}$). This superior performance extends up to about 800 km during nighttime operations. Furthermore, during daylight, even though the TDS C-band tolerates lower $\theta_\mathrm{elev}^\mathrm{max}$ angles than others, SPS H\textalpha\ and Ca~II continue to outperform the TDS C-band up to 750 km and 950 km ground track offset, respectively. This SPS advantage is clearly displayed when we estimate the key length volume over a year.\\
\indent Table \ref{tab:Annual_SKL} presents $\scriptstyle \overline{\mathrm{SKL}}^{\, \mathrm{equator}}_{\, \mathrm{annual}}$, the expected annual secret key length, using Eqs.\ \ref{eqn:SKLannualDayNight} and \ref{eqn:SKLannual}, for under investigation wavelengths and sources during daylight and night time operation assuming the ground station at the equator. To obtain those SKLs we integrate the SKL in Fig.\ \ref{fig:SKL_track_offset} using Eq. \ref{eqn:SKL_int}. The corresponding outcomes in bits-meter unit are given in Table SII. Operating a TDS at night, H\textalpha\ shows higher annual SKL than Ca~II and C-band, while during daylight Ca~II outperforms the other two operating wavelengths. Considering an SPS, the H\textalpha\ line is superior to the other two candidates during daylight and night time operations. Overall, we estimate that SPS H\textalpha\ generates the highest SKL of 7.5763 Gb per year among all sources.\\
\indent In general, the H\textalpha\ line has the lowest system loss metric $\eta_\mathrm{sys}^\mathrm{loss}$, i.e., dB loss of the quantum link at the zenith, amongst all wavelengths, while C-band clearly starts with a considerable disadvantage compared to the visible and infrared candidates, see the $\eta_\mathrm{sys}^\mathrm{loss}$ in Table \ref{tab:Annual_SKL}. This gap of approximately 6 and 5 dB loss in relation to H\textalpha\ and Ca~II, respectively, is mainly due to a higher diffraction loss of the C-band, since longer wavelengths experience higher beam broadening (see Eq. \ref{wG}) and consequently the transmission efficiency through the ground station telescope is lower. Albeit the channel loss difference between theses wavelengths decreases during a satellite pass, the secret key length for C-band does not exceed the other wavelengths in most scenarios. Even if the channel loss assumptions made in this study might seem idealistic when contrasted with the Micius satellite implementation, they remain within the realm of feasibility given the constant advancement in satellite communication technologies, targeting lower losses. However, it is worth noting that in scenarios with higher losses, a TDS could potentially outperform a realistic SPS. This conclusion can be drawn from Fig.\ \ref{fig:SKL_altitude} and Fig.\ \ref{fig:SKL_track_offset}. Both figures illustrate that TDS performs better in the presence of very high losses, whether these losses are due to increased altitudes or larger ground track offsets.\\
\indent In the event of ground station telescopes being coupled to (short) optical fibers, our findings concerning the optimal operating wavelength remain applicable. Any additional system loss incurred due to coupling efficiencies can be relatively wavelength-independent, thus, not altering the relevance of the ideal operating wavelength identified in this study.\\
\section{\label{sec:level1}Proposal of a suitable single-photon source}
In the simulations conducted, a single-photon source with the performance analogous to what we have demonstrated using a cavity-enhanced quantum emitter \cite{vogl2019compact,PhysRevResearch.3.013296} surpasses TDS. The room-temperature SPS is based on a fluorescent defect hosted by hBN, coupled to a confocal microcavity (consisting of a flat and a hemispherical dielectric mirror). Subsequently, we propose a method by which the demonstrated photon source can be adapted to implement the daylight QKD scheme at a Fraunhofer line. We must keep in mind that integrating the source into the satellite for the downlink scenario imposes limitations on its size, weight, and power requirements for operation. Our source, however, has been implemented on a satellite prototype within a volume of 1 l \cite{vogl2019compact}, therefore already fulfilling these requirements.\\
\indent Further prerequisites are emission at a Fraunhofer line and tunability (in case the fabrication accuracy is less than a few 10s of pm). The linewidth should be below 0.5 nm (which is the assumed spectral filtering width above). With the demonstrated emission linewidth of 0.2 nm (FWHM), the source also already meets this requirement. The cavity funnels the emission into the resonant mode and thereby narrows down the spectrum. The central wavelength can be controlled with a piezo actuator. For locking the cavity to the specific Fraunhofer line, a feedback loop is required. It is possible to route a fraction of the cavity-emitted photons to a reference gas cell containing the element responsible for the Fraunhofer line. The cavity length (i.e., resonance wavelength) will then be varied to minimize transmission through this reference gas cell detected by a single-photon detector after the cell. By that means the absorption of the gas is maximized, and in this case, the cavity emits in resonance with the atomic transition.\\
\indent This leaves the final question of which color center could be used. For defects in hBN, which have been used for QKD already \cite{10.1002/qute.202200059,https://doi.org/10.1002/qute.202300038}, a large distribution of color centers has been experimentally demonstrated \cite{C9NR04269E,doi:10.1021/acsnano.6b03602,PhysRevB.98.081414}. In theoretical simulations, defects at specific Fraunhofer lines have been predicted \cite{cholsuk2022tailoring}. Overall, hBN is better suited compared to other solid-state quantum emitter systems, due to the broad coverage of emission lines in general. These lines, however, will likely not completely overlap with the Fraunhofer lines. 2D materials are also better in this respect, as residual strain in a 2D material can shift the transition lines, typically by much more than is possible in 3D crystals (such as diamond) \cite{cholsuk2022tailoring}. For the H\textalpha\ line (656 nm), the P$_\text{N}$V$_\text{B}$ defect in hBN (where a phosphor atom replaces a nitrogen atom next to a nitrogen-vacancy) is a suitable candidate. Considering Ca~II line (854 nm), the Er$_\text{N}$V$_\text{N}$ defect in hBN (where an erbium atom replaces a nitrogen atom next to a nitrogen-vacancy) is a proper candidate. Such defects could be fabricated by ion implantation and then integrated into a cavity with a locking scheme as described above. Moreover, hBN quantum emitters have also been qualified for use in space environments \cite{vogl2019radiation}.
\section{\label{sec:conc}Conclusion}
In this work, we have calculated the ideal wavelength for satellite-based QKD during daylight conditions. We have modeled realistic parameters for the sender and receiver units based on the performance of commercially available components. Moreover, we have also included different light source types in our analysis and considered the finite-key effect. Our results show that for achieving the maximum secret key rate during daylight, the H\textalpha\ Fraunhofer line at 656 nm is the ideal wavelength, and a realistic SPS can provide more than 10 time higher annual secret key length than any source operating at C-band. Moreover, we find that realistic state-of-the-art single-photon sources are already good enough to outperform decoy protocols. In contrast to contemporary daylight QKD systems in the near-infrared (where the solar background is low during the day), visible wavelengths turn out to be more efficient when all factors are considered. We have also proposed a realistic scheme for a single-photon source that can implement daylight QKD at a Fraunhofer line.\\
\indent As we provide the complete model, our calculations can easily be adapted for other scenarios as well. In particular, our results can also be generalized to terrestrial roof-to-roof quantum links. Our work therefore will have applications for the realization of a future quantum internet that features high data rates that are operational at any time.
\section*{Data availability}
The data generated during this work are available from the corresponding author T.V. upon reasonable request.
\section*{Notes}
T.V. declares to have filed a patent for a single photon source that can implement such daylight QKD scheme.\\
\\
\begin{acknowledgments}
This work was funded by the Deutsche Forschungsgemeinschaft (DFG, German Research Foundation) - Projektnummer 445275953. The authors acknowledge support by the German Space Agency DLR with funds provided by the Federal Ministry for Economic Affairs and Climate Action BMWK under grant numbers 50WM2165 (QUICK3) and 50RP2200 (QuVeKS). T.V. is funded by the Federal Ministry of Education and Research (BMBF) under grant number 13N16292. C.C. acknowledges a Development and Promotion of Science and Technology Talents Project (DPST) scholarship by the Royal Thai Government. D.K.L.O. and R.G.P. acknowledge support from the UK NQTP, the EPSRC Quantum Technology Hub in Quantum Communications (EP/T001011/1), and the EPSRC International Network in Space Quantum Technologies (EP/W027011/1). R.G.P. is supported by the EPSRC Research Excellence Award (REA) Studentship and acknowledges the SUPA Saltire Emerging Research Visits scheme. D.K.L.O. is an EPSRC Researcher in Residence at the Satellite Applications Catapult (EP/T517288/1).\\
\end{acknowledgments}

\end{document}